\documentclass[a4paper]{VisionStyle}
\usepackage{epsfig}

\begin{document}

\title{A High Resolution X--ray spectrum of the LINER Galaxy M81}

\author{M.J.\,Page\inst{1} \and A.A.\,Breeveld\inst{1} \and 
  R.\,Soria\inst{1} \and K.\,Wu\inst{1}} 

\institute{
  Mullard Space Science Laboratory, University College London, Holmbury
  St. Mary, Dorking, Surrey RH5 6NT, UK
}

\maketitle 

\begin{abstract}
We present the soft X--ray spectrum of the LINER galaxy M81 from a
long observation with the {\em XMM-Newton} RGS. As well as continuum emission
from the active nucleus, the spectrum shows emission lines from
multi temperature collisionally ionized gas. The
emission lines are significantly broader than the RGS point-source
spectral resolution, and in the cross dispersion
direction the emission lines are detected adjacent to, as well as
coincident with, the active nucleus. This implies that they
originate in a region of more than an arcminute ($\sim 1$ kpc) spatial
extent. A good fit to the whole RGS spectrum can be found using a model
consisting of an absorbed power law from the active nucleus and 
a 3 temperature thermal plasma. Two of the thermal plasma
components have temperatures of 0.18$\pm 0.04$ keV and 0.64$\pm 0.04$ keV, 
characteristic of the 
hot interstellar medium produced by supernovae; 
the combined luminosity of these two plasma components
accounts for all the unresolved bulge X--ray emission seen in the 
{\em Chandra} observation by \cite*{mpage-C2:tennant01}. The third component
has a higher temperature ($1.7^{+2.1}_{-0.5}$ keV) and we propose
X--ray binaries in the bulge of M81 as the most likely source of this emission.

\end{abstract}

\section{Introduction}
\label{sec:introduction}

M81 is an Sab spiral galaxy hosting a low luminosity Seyfert nucleus
which shows the characteristics of a ``low ionization nuclear
emission-line region'' (LINER, \cite{mpage-C2:heckman80}). 
It was first
observed in X--rays with {\em Einstein} 
(\cite{mpage-C2:elvis82}, \cite{mpage-C2:fabbiano88}) 
which resolved several discrete sources in M81, the
brightest of which is associated with its active nucleus. An
apparently diffuse emission component associated with the bulge of
M81, spatially extended over a few arcminutes was detected first with
{\em Rosat} (\cite{mpage-C2:roberts00}, 
\cite{mpage-C2:immler01}) and later confirmed with Chandra data 
(\cite{mpage-C2:tennant01}). Spectra from 
{\em Rosat} (\cite{mpage-C2:immler01}), {\em BBXRT}
(\cite{mpage-C2:petre93}), {\em ASCA} (\cite{mpage-C2:ishisaki96}) and
{\em BeppoSAX} (\cite{mpage-C2:pellegrini00}) suggested the presence of
a soft ($<$ 1 keV) thermal component in addition to an absorbed power law
component from the nucleus. In this paper we present the soft X--ray
spectrum of M81 at much higher resolution, from an observation with
the {\em XMM-Newton} reflection grating spectrometer (RGS, 
\cite{mpage-C2:denherder01}).

\section{RGS data and spectral analysis}
\label{mpage-C2_sec:observation}

M81 was observed by {\em XMM-Newton} in April 2001
for a total of 
138 ks. 
The RGS data were reduced using the {\em XMM-Newton} standard analysis 
software (SAS) version 5.2.
The source spectrum was
extracted from a region centred on the nucleus, enclosing 90\% of the point spread function in the
cross dispersion direction. Background regions were selected in the
cross-dispersion direction so as to exclude 99\% of the nuclear point spread
function.

\begin{figure*}
  \begin{center}
    \epsfig{file=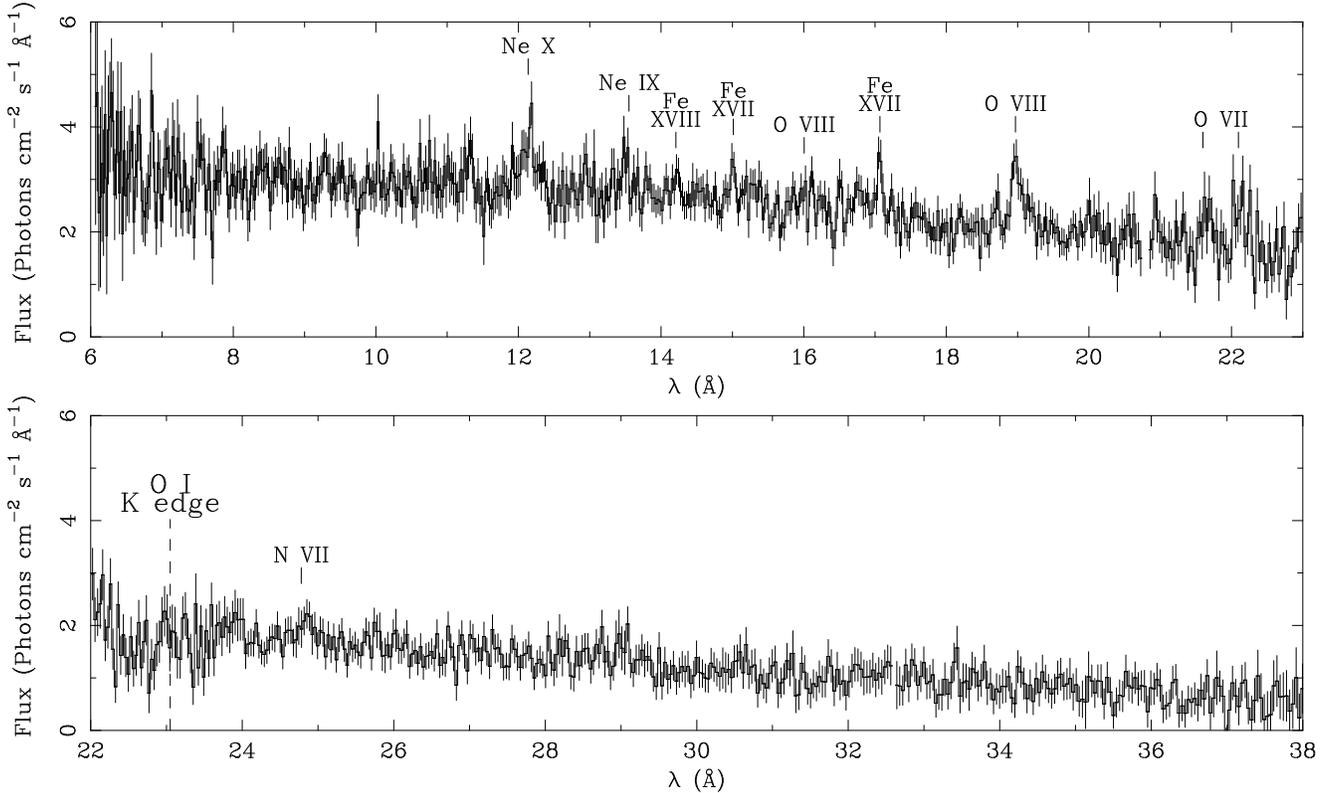, width=10.5cm, angle=270}
  \end{center}
\caption{The RGS spectrum with prominent emission lines marked}
\label{mpage-C2_fig:rgsspectrum}
\end{figure*}

\subsection{X-ray emission line gas}
\label{mpage-C2_sec:emissionlines}

Figure~\ref{mpage-C2_fig:rgsspectrum} shows the RGS spectrum. A number of prominent 
emission lines from H-like and He-like N, O, and Ne as well as L shell lines
from Fe XVII and Fe XVIII are visible above the continuum; 
these are labelled in Figure~\ref{mpage-C2_fig:rgsspectrum}. 
The lines are considerably broader
 than the RGS line spread function
for a
point source. For example, O VIII Ly$\alpha$ at $\sim 19$ \AA\ is expected to 
have relatively little contamination from
neigbouring lines, but a gaussian fit to this line shows that it is 
inconsistent with $\sim \sigma = 0$ at $> 99\%$.
This broadening could be due to kinematic motions in the gas, for example 
if it is associated with the active nucleus. However 
a similar level of line broadening
would result from the gas being extended over a region of $\sim$ 2 arcminutes, 
because the RGS is a slitless dispersive spectrograph and its resolution is
degraded for extended objects. 

\begin{figure}
  \begin{center}
    \epsfig{file=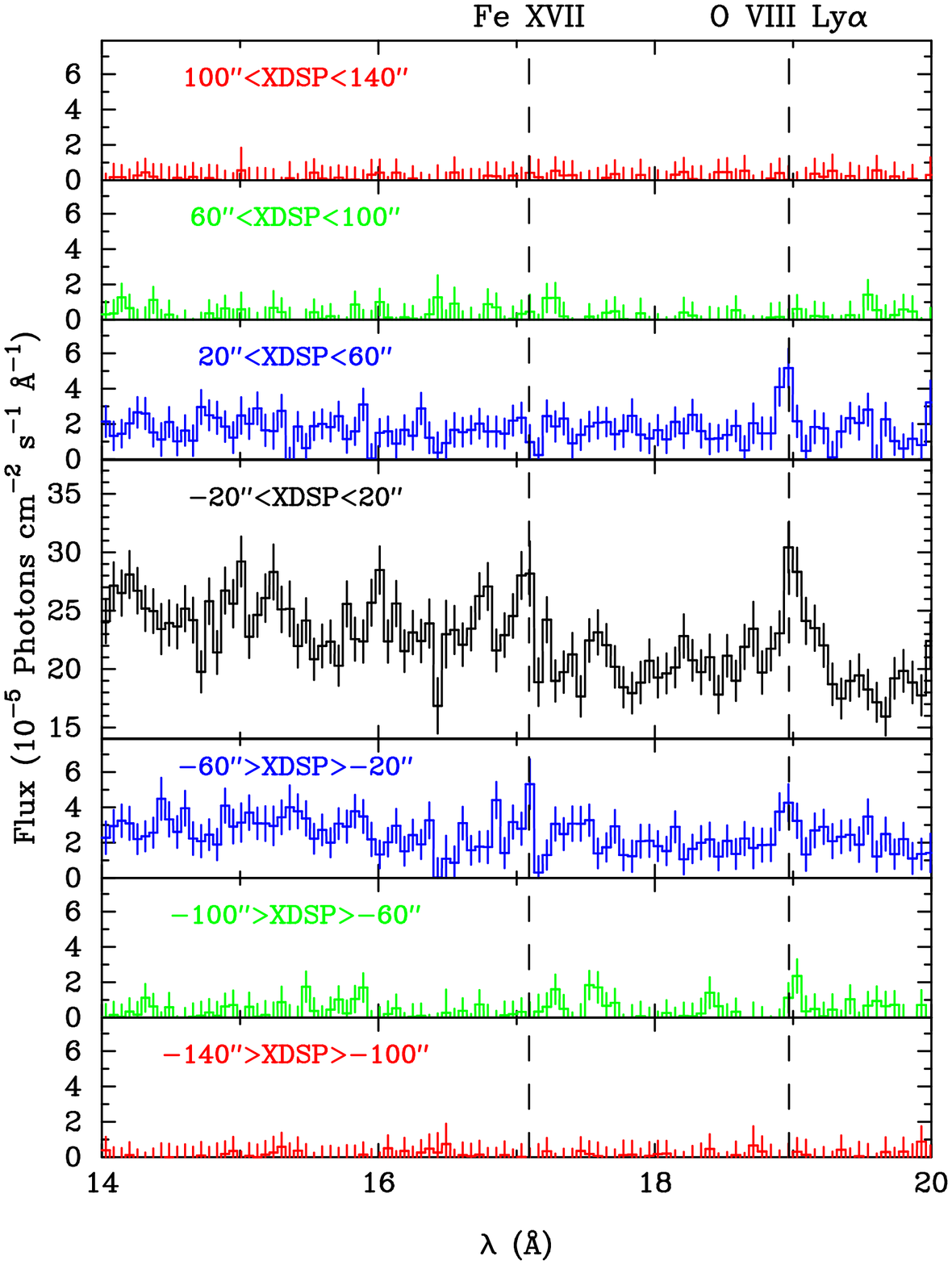, width=8.5cm}
  \end{center}
\caption{Spectral cuts in the cross dispersion direction showing that the
 line emission comes from a region that is extended in the cross dispersion 
direction}
\label{mpage-C2_fig:crosscuts}
\end{figure}

To distinguish these alternatives, we have examined the spatial
distribution of the emission line gas in the RGS cross dispersion direction by
constructing spectra in $40\arcsec$ wide strips at a range of distances from
the nucleus of M81. A background
spectrum was constructed from the average of the 
two most off-axis strips. The
emission-line rich 14-20 \AA\ parts of these spectra are shown in
Figure~\ref{mpage-C2_fig:crosscuts}. The emission line gas 
is extended over more than an arcminute implying that the excess
broadening of the emission lines is due to their spatial extent rather than
kinematic motions.

\begin{figure}
  \begin{center}
    \epsfig{file=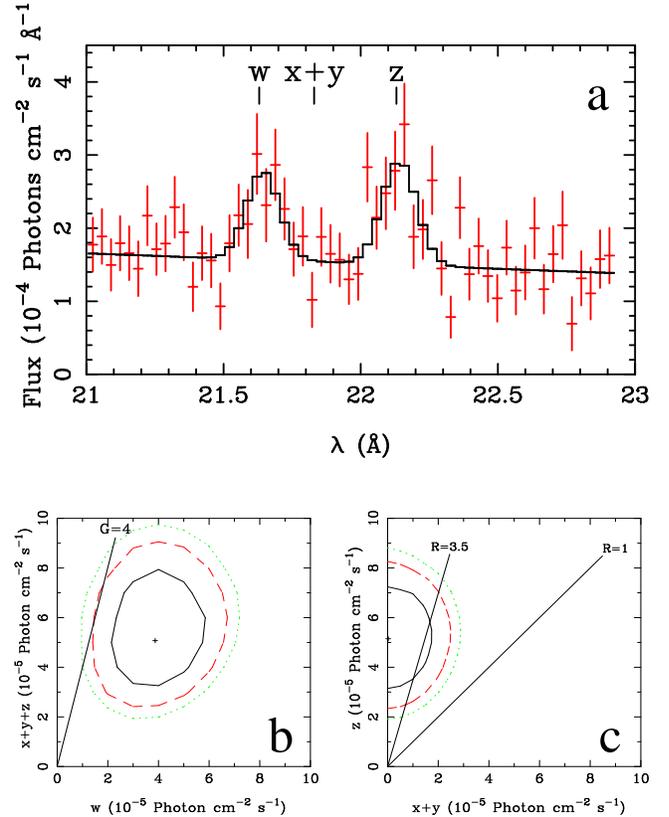, width=8.5cm}
  \end{center}
\caption{{\bf a)} The O VII triplet and best fit model\ \ {\bf b)} confidence
contours (68\%, 90\% and 95\%) for the strength of the OVII x+y+z lines 
against the w line strength\ \
{\bf c)} confidence contours for the OVII z line strength against the x+y 
lines; see Section \ref{mpage-C2_sec:emissionlines}}
\label{mpage-C2_fig:oxygenplots}
\end{figure}

We further investigate the nature of the emission line gas using plasma
diagnositics from the He-like OVII triplet, as described in
\cite{mpage-C2:pourquet00}. 
Taking the 21-23 \AA\ portion of the RGS spectrum, 
we fitted the 
OVII lines as 3 gaussians with the relative wavelengths of the resonance (w),
intercombination (x+y) and forbidden (z) lines fixed at their theoretical
ratios. The best fit has an acceptable $\chi^{2}/\nu = 56/51$, and is shown
superimposed on the 21-23 \AA\ RGS spectrum in Figure~\ref{mpage-C2_fig:oxygenplots}a. 
In Figure~\ref{mpage-C2_fig:oxygenplots}b 
we show the confidence contour of the (x+y+z)
against w line strength. 
The solid line shows the line ratio G 
(=(x+y+z)/w) = 4. 
Plasmas in which photoionization is the dominant ionization mechanism are
expected to have ratios to the left of this line (i.e. G $>$ 4); 
this is ruled out at 95\% confidence in M81. 
In plasmas with line ratios to the right of this line, such
as that seen in M81, collisional excitation
is important and
may be the dominant emission mechanism.
Figure~\ref{mpage-C2_fig:oxygenplots}c 
shows confidence contours of the density diagnostic z
against (x+y). The line shown on the plot corresponding to ratio R
(=z/(x+y)) = 1 is excluded at $> 95\%$ along with all of parameter space to the
right of this line, implying that the line emitting plasma has a density of $
{\rm n}_{e} <
10^{11} {\rm cm^{-3}}$. 

\begin{table}
\caption{Spectral fitting of the whole M81 RGS spectrum}
\label{mpage-C2_tab:totalfits}
\begin{tabular}{lcccc}
(1)&(2)&(3)&(4)&(5)\\
Model&$\Gamma$ or $kT$&flux&${\rm N_{H}}$&$\chi^{2}/\nu$\\
\hline
&&&&\\
PL$\times{\rm N_{H}}$        &$2.06\pm0.06$&$99.9\pm 3.8$&$4.8\pm0.8$&1085/992\\ 
&&&&\\
PL$\times{\rm N_{H}}$     &$1.94\pm0.07$&$95.8\pm 3.7$&$3.3\pm0.8$&876/990\\ 
+Mek &$0.66\pm0.04$&$4.3^{+0.9}_{-0.7}$&&\\ 
&&&&\\
PL$\times{\rm N_{H}}$    &$1.93\pm0.07$&$94.5\pm3.8$&$3.4\pm0.8$&821/984\\ 
+Mek3 &$0.20^{+0.04}_{-0.02}$&$1.5\pm 0.5$&&\\ 
+Mek3 &$0.66\pm0.05$&$4.2^{+0.8}_{-0.7}$&&\\ 
&&&&\\
PL$\times{\rm N_{H}}$    &$1.94\pm0.06$&$91.6^{+7.4}_{-4.4}$&$3.4\pm0.8$&803/980\\ 
+Mek3 &$0.18\pm0.04$&$1.1^{+0.5}_{-0.4}$&&\\ 
+Mek3 &$0.64\pm0.04$&$4.2^{+0.8}_{-0.9}$&&\\ 
+Mek3 &$1.7^{+2.1}_{-0.5}$&$3.4^{+7.6}_{-2.2}$&&\\ 
&&&&\\
\hline
&&&&\\
\multicolumn{5}{l}{Explanation of columns:}\\
\multicolumn{5}{l}{1 PL = power law,}\\ 
\multicolumn{5}{l}{\ \ NH = neutral absorber (`tbabs'
model in {\scriptsize XSPEC}),}\\
\multicolumn{5}{l}{\ \ Mek = thermal plasma (`Mekal' model in {\scriptsize
XSPEC}),}\\
\multicolumn{5}{l}{\ \  Mek3 = thermal plasma
(`Mekal' model in
{\scriptsize XSPEC}) with}\\
\multicolumn{5}{l}{\ \  components
at $\pm 1500$ km s$^{-1}$ to the broadening of }\\
\multicolumn{5}{l}{\ \  the lines produced by the
spatial extent of the emission.}\\ 
\multicolumn{5}{l}{2 power law photon index or temperature of thermal}\\
\multicolumn{5}{l}{\ \ plasma (keV)}\\
\multicolumn{5}{l}{3 flux of model component in the energy range 0.3-2.0 keV,}\\
\multicolumn{5}{l}{\ \ in units of $10^{-13}$ erg cm$^{-2}$ s$^{-1}$
after correction for}\\
\multicolumn{5}{l}{\ \ Galactic absorption 
of ${\rm N_{H}} = 4.16\times 10^{20}$ cm$^{-2}$}\\
\multicolumn{5}{l}{4 intrinsic neutral column density (cm$^{-2}$).}\\
\end{tabular}
\end{table}

\subsection{Modelling the whole RGS spectrum}
\label{mpage-C2_sec:modelling}

The 
emission lines in the
soft X--ray spectrum of M81 are superimposed on a strong continuum
from the its active nucleus. 
To reproduce this continuum emission we used {\small XSPEC} to fit
 a power law model,
modified by both Galactic column and intrinsic absorption ({\small xspec}
 model TBabs of
\cite{mpage-C2:wilms00}). 
We obtained a ($\chi^{2}/\nu = 1085/992$) for an intrinsic
absorber with a
column
density of  ${\rm N_{H}}\sim$ 5 $\times
10^{20}$.
The fit parameters 
of this fit, and subsequent fits, are given in
Table \ref{mpage-C2_tab:totalfits}.

\begin{figure*}
  \begin{center}
    \epsfig{file=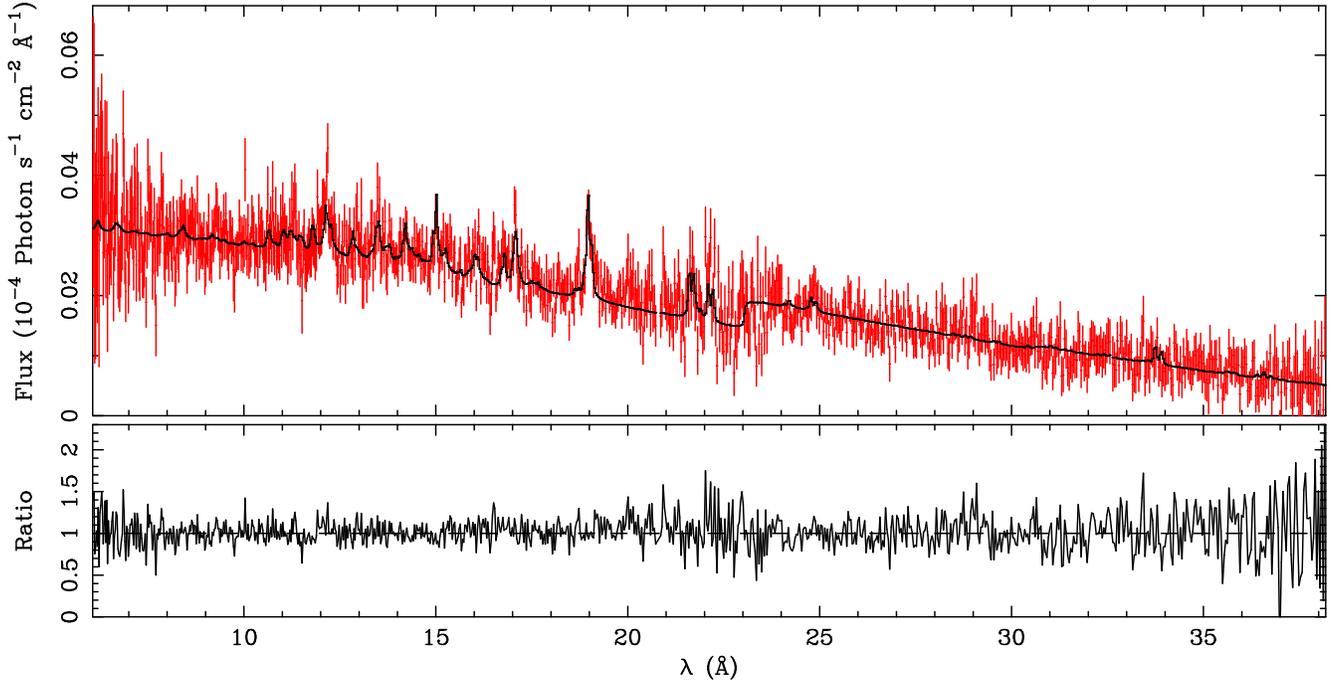, width=9cm, angle=270}
  \end{center}
\caption{Power law + intrinsic neutral absorption + 3 mekal model fit to the
RGS spectrum. The top panel shows the model (black line) and the data (red
points). The bottom panel shows the data/model ratio}
\label{mpage-C2_fig:rgsfit}
\end{figure*}

We next attempted to model the emission line component. As
shown in Section \ref{mpage-C2_sec:emissionlines} the O VII triplet is 
consistent with
an origin in a collisionally ionized gas. 
We therefore began modelling the emission
lines by adding a single temperature `Mekal' thermal plasma to the
absorbed power law model. This
resulted in a significant improvement in the fit ($\chi^{2}/\nu = 876/990$)  
for a
plasma temperature of $0.66\pm 0.04$ keV. However, the emission lines are too
narrow in this model, and some 
emission lines,
most noticably those from OVII, have much lower intensities in the 
 model than in the observed spectrum. We therefore added a second, lower
temperature, Mekal component to the model, and at both Mekal temperatures we
added components with positive and negative velocities of 1500 km s$^{-1}$ to 
reproduce the width of the lines caused by the spatial extent of the emission 
region.
 The relative normalisations of the rest,
+1500 km s$^{-1}$ and -1500 km s$^{-1}$ components were allowed to vary freely,
but the temperatures were tied in the fit; the fluxes given for these model
components in Table \ref{mpage-C2_tab:totalfits} are the totals for all three 
`velocity' 
components. This resulted in a significant improvement in $\chi^{2}$,
reproducing most of the emission lines well, with the second plasma component
at a best fit temperature of $\sim 0.2$ keV. However, the blend of Ne X with Fe
XXII-XXIII at $\sim 12.1$ \AA\ is still underproduced by this model, and so a
third, higher temperature, Mekal plasma was
added. This resulted in another significant improvement in the fit, to a 
final $\chi^{2}/\nu = 803/980$, with a best fit temperature for the third 
Mekal component of $1.7^{+2.1}_{-0.5}$ keV. The model reproduces the spectrum
well, as shown in Figure~\ref{mpage-C2_fig:rgsfit}.

\section{Discussion}

In Section \ref{mpage-C2_sec:emissionlines} we showed that the soft X--ray 
emission lines in M81 come from an extended region. 
We also showed that the diagnostic OVII triplet is emitted by a low
density region and is not powered by photoionization from the central AGN.
Furthermore, in the previous section we showed that 
a 3 temperature thermal plasma model provides
a good description of the emission line component of the M81 RGS spectrum in
conjunction with absorbed powerlaw emission from the nucleus. 
Two of the thermal plasma components 
have temperatures of $0.18\pm0.04$ keV and $0.64\pm0.04$ keV, 
which are too low to be produced by accreting binaries, and
are instead more characteristic of the hot interstellar medium produced by 
supernova explosions; 
in this case we would expect these emission lines to come 
from a genuinely diffuse region (or regions) in the bulge of M81. 

The X--ray imaging study performed with {\em Chandra} does indeed find 
diffuse emission from the bulge of M81. 
 The luminosity of this unresolved bulge
emission reported by \cite*{mpage-C2:tennant01} is $8.6 \times 10^{38}$
erg s$^{-1}$ (0.2-8 keV) assuming a distance of 3.6 Mpc to M81. Assuming the
same distance, the 0.18 keV and 0.64 keV thermal plasma components from our
best fit model in Table 1 have luminosities of $2.2 \times 10^{38}$ erg
s$^{-1}$ and $6.9 \times 10^{38}$ erg s$^{-1}$ respectively over the same
energy range - sufficient to completely account for the unresolved bulge 
emission.

If the emission associated with the highest of the three plasma temperatures 
($1.7^{+2.1}_{-0.5}$ keV) were also produced by a hot interstellar medium, or
by a large population of very low luminosity sources such as cataclysmic
variables, it would also appear diffuse in the {\em Chandra} imaging. 
However, the 
two lower temperature plasma components alone completely account for the
unresolved emission, and hence the
higher temperature emission must arise in part of the bulge population resolved
by {\em Chandra}. We propose X-ray binaries as the most likely origin of the
higher temperature emission.

\begin{acknowledgements}

Based on observations obtained with {\em XMM-Newton}, an ESA science mission
with instruments and contributions funded by ESA Member States and the
USA (NASA).

\end{acknowledgements}

\end{document}